\documentclass[10pt, conference, compsocconf]{IEEEtran}

\usepackage{cite}

\usepackage{graphicx}

\usepackage{amssymb}
\usepackage{amsmath}

\usepackage{multirow}

\usepackage{subfigure}

\usepackage[hyphens]{url}
\usepackage[bookmarks=false]{hyperref}
\usepackage{breakurl}

\usepackage{lipsum}
\usepackage{pdfpages}

\newcommand\blfootnote[1]{%
  \begingroup
  \renewcommand\thefootnote{}\footnote{#1}%
  \addtocounter{footnote}{-1}%
  \endgroup
}

\usepackage{./AMMALanguages}

\newcommand{\Keywords}{\lstset{keywords={provided,is,within,and,or,(,),TRUE,FALSE,tr-policy}}}
\newcommand{\code}[1]{{\footnotesize{\texttt{#1}}}}

\pdfoutput=1

\begin{document}

\mathchardef\mhyphen="2D

\includegraphics[trim=30mm 120mm 30mm 30mm,clip,width=\textwidth]{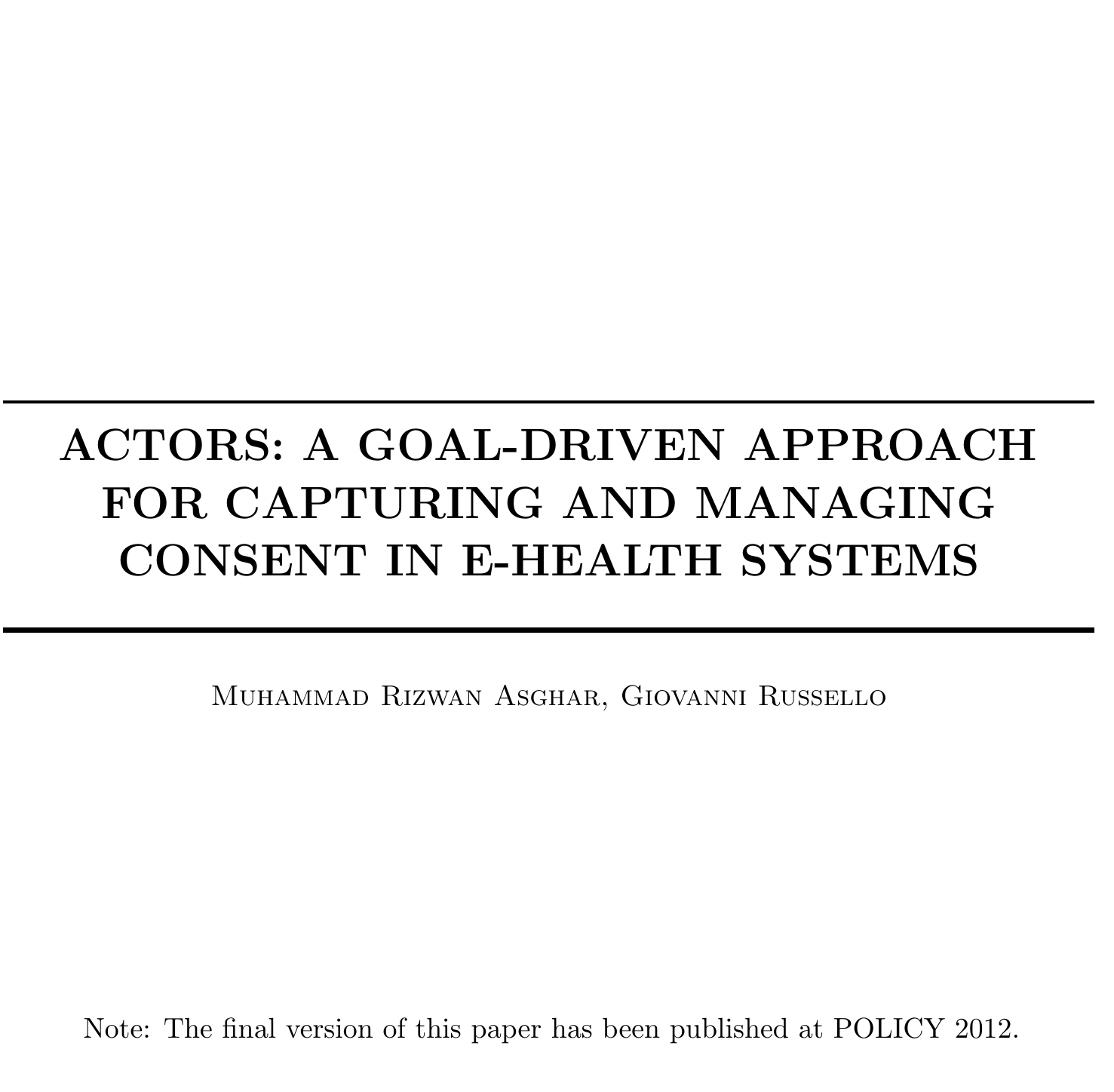}

\pagestyle{plain}
\pagenumbering{arabic}

\title{ACTORS: A Goal-driven Approach for Capturing and Managing Consent in e-Health Systems}

\author{
    \IEEEauthorblockN{Muhammad Rizwan Asghar\IEEEauthorrefmark{2}\IEEEauthorrefmark{3}, Giovanni Russello\IEEEauthorrefmark{4}}
    \IEEEauthorblockA{\IEEEauthorrefmark{2} CREATE-NET, International Research Center, Trento Italy
    \\asghar@create-net.org}
    \IEEEauthorblockA{\IEEEauthorrefmark{3} Department of Information Engineering and Computer Science, University of Trento, Trento Italy
    \\asghar@disi.unitn.it}
    \IEEEauthorblockA{\IEEEauthorrefmark{4} Department of Computer Science, The University of Auckland, Auckland New Zealand
    \\g.russello@auckland.ac.nz}
}

\maketitle

\blfootnote{* The final version of this paper has been published at POLICY 2012 \cite{Asghar2012-POLICY}.}

\begin{abstract}

The notion of patient's consent plays a major role in granting access to medical data. In typical healthcare systems, consent is captured by a form that the patient has to fill in and sign. In e-Health systems, the paper-form consent is being replaced by the integration of the notion of consent in the mechanisms that regulate the access to the medical data. This helps in empowering the patient with the capability of granting and revoking consent in a more effective manner. However, the process of granting and revoking consent greatly varies according to the situation in which the patient is. Our main argument is that such a level of detail is very difficult and error-prone to capture as a set of authorisation policies. In this paper, we present ACTORS, a goal-driven approach to manage consent. The main idea behind ACTORS is to leverage the goal-driven approach of Teleo-Reactive (TR) programming for managing consent that takes into account changes regarding the domains and contexts in which the patient is providing her consent.

\end{abstract}


\section{Introduction}
Healthcare information refers to any data containing information about an individual's health conditions. As it contains sensitive personal information, its improper disclosure may influence several aspects of an individual's life. Today, medical data is massively being converted into electronic format. Individuals' medical data can be now easily accessible to a very large number of health-care professionals. Although this is done with the best of intentions to improve the processing and streamline healthcare delivery, it also poses very concrete threats to the individual's privacy.

Since the medical information of an individual is confidential, the only basis for accessing it is through that individual's consent \cite{EC1995}. In traditional healthcare systems, an individual provided her consent by signing a paper form. In these settings, withdrawing consent was very difficult for an individual because she had to go through complicated bureaucratic processes. Moreover, the granularity of consent was very coarse-grained. The individual agreed in providing consent in advance for all her medical data, thus violating the principle of least privilege.

Policy-based authorisation mechanisms have successfully been used in managing access rights given the flexibility and re-usability that they offer. In literature, several approaches have been realised where the notion of consent is integrated with the policy decision mechanism. For instance, Russello \emph{et al.} \cite{Russello2008} propose to capture the notion of consent through the use of medical workflows and to integrate it with Ponder2 authorisation policies\footnote{\url{http://ponder2.net/}}. Wuyts \emph{et al.} \cite{Wuyts2011} have extended the XACML \cite{OASIS} authorisation model with the notion of consent.


To specify a set of authorisation policies that capture all the details required to enforce correctly an individual's decisions about consent is very complex. First of all, each authorisation policy has conditions to express when it should be enforced that might be in conflict with other policies. Although work has been done to address the problem of automatically resolving conflicts \cite{Russello2007}, it is not possible to completely automate the decision since in the specific case of the healthcare scenario humans are also involved. To complicate matters further, contextual information needs to be captured to identify the purpose of the access being requested. If these details are not captured correctly in the policy specification by the security administrator then there may be serious consequences.

For instance, the way in which an individual wants to provide and revoke her consent differs according to the caregivers that she is interacting with. With her General Practitioner (GP), a patient typically establishes a lasting relationship; therefore, consent can be given for a long time. On the other hand, when she is visiting a specialist in a hospital, she wants to give consent only for the time the treatment will last and only for the data that is required for the specific treatment. Still, another different situation is in the case of an emergency where the paramedics have to provide first care before reaching the emergency room. In this case, consent can be given to the medical data however for the short period of time required to reach the hospital.

From the above scenario, it emerges that specifying in one single policy set all the requirements for managing consent is a very error-prone task. In the light of this, in this paper we propose ACTORS (Automatic Creation and lifecycle managemenT Of authoRisation policieS) where a goal-driven approach is used to \emph{glue} together and manage authorisation policies that have a common aim, that is the handling of consent in a specific context (i.e., consent for the GP, for the specialist, and paramedics). In particular, our observation is that we can simplify the specification of authorisation policies when these are treated as a \emph{program sequence} towards a specific goal. The main contribution and novelty of this paper is to propose the idea of using Teleo-Reactive (TR) programs to glue together authorisation policies aiming at a specific goal. The idea of TR programs was initially introduced by Nilsson \cite{Nilsson1994}. The main advantage of TR programs is that the way in which they are specified is very natural for humans. Therefore, a security administrator can capture more naturally the security requirements in a TR sequence.

The rest of this paper is organised as follows. In Section \ref{sec:related_work}, we review the related work. Section \ref{sec:case_study} provides an overview of a case study that we use to demonstrate the feasibility of our approach. Next, we provide a brief overview of Teleo-Reactive Policies in Section \ref{sec:tr-ovewview}. In Section \ref{sec:proposed_approach}, we present our proposed approach. In Section \ref{sec:application}, we present how the case study scenarios can be modelled using the proposed approach. Finally, we conclude in Section \ref{sec:conclusion_future_work} and indicate the direction of our future work.

\section{Related Work}
\label{sec:related_work}


Marinovic \emph{et al.} \cite{Marinovic2010} employ TR policies for continuously monitoring the nursing home, where caregivers (including nurses, head-nurses, patients and students) are equipped with mobile devices for running their corresponding TR policies. They use TR policies to manage all activities of a caregiver using one workflow specification while we use TR policies with the goal of capturing consent that may involve instantiation of authorisation policies regarding consent and management of their lifecycle, consisting withdrawal and activation of consent.

Illner \emph{et al.} \cite{Illner2005-PDCN, Illner2005-INDIN} suggest an automated approach for managing services related to distributed and embedded systems in dynamic environments. In their approach, various configurations for the services are generated and mapped to specific environmental conditions only once at the design time when system is setup while appropriate configurations for the services are activated at runtime when certain environmental conditions hold. The shortcoming of this approach is that the configurations are defined statically while our goal-based approach is dynamic in a sense that authorisation policies do not need to be specified in advance and are instantiated automatically while taking into account environmental conditions.



Johnson \emph{et al.} \cite{Johnson2010} suggest a general approach for creating policy templates. A policy template provides users with a structured format for authoring policies. In our proposed solution, a healthcare provider may consider this work for generating policy templates. Chan and Kwok \cite{Chan2006} describe a method to create policies automatically based on observed events. They use the Singular Value Decomposition (SVD) technique for modelling correlation between events and policies and then create new policies or select recommended policies based on the correlation. Unfortunately, the SVD technique may not always choose the fine-grained policies while our proposed approach always generates the fine-grained authorisation policies based on environmental conditions.

Fu \emph{et al.} \cite{Fu2001, Fu2001-phdthesis} propose how to automatically generate required IPSec policies without manual configuration. The idea is to define high-level security requirements and then automatically generate a set of IPSec policies that can satisfy all security requirements. The main problem is that this approach incurs high performance overhead for finding the required set of policies as the proposed algorithm needs to go through a large number of possibilities before halting. Instead of generating a set of authorisation policies, our proposed approach generates only a single authorisation policy while taking into account contextual information and user intent.


Russello \emph{et al.} \cite{Russello2008} propose a consent-based framework that enables patients to control disclosure of their medical data, where the mechanism of capturing consent is integrated with workflows. The idea is to automatically generate Ponder2 style of authorisation policies \cite{Damianou2001} that depend on workflows. However, there is no automatic mechanism for managing the lifecycle of consent, such as consent withdrawal, activation or deletion. Asghar and Russello \cite{Asghar2011} suggest a mechanism for managing the consent lifecycle. They introduce a notion of very expressive consent represented as a consent policy. However, they assume that a data subject defines his/her consent policies; unfortunately, such a solution may not be acceptable because data subjects may not be able to understand low-level policy details.

Wuyts \emph{et al.} \cite{Wuyts2011} incorporate patient consent with healthcare systems. They use the XACML policy language \cite{Moses2003} (proposed by OASIS \cite{OASIS}) for defining access control on medical data and retrieve consent from the Policy Information Point (PIP). They express consent as a set of pre-defined attributes and store it in the database. The similar approach is used by Jin \emph{et al.} in \cite{Jin2009}, which is an authorisation framework for sharing Electronic Health Record (EHR). The main issue with both approaches is that the set of pre-defined attributes may not be sufficient to capture consent as it may involve certain conditions. In order to overcome this issue, there are approaches \cite{Asghar2011, OKeefe2005} in which consent is treated as an authorisation policy; however, it raises some other problems. First, this approach requires users to specify low-level details, which a normal user may not be aware of, at the time of policy creation. Second, there is no automatic mechanism for managing the consent lifecycle.

EnCoRe \cite{encore}, a currently ongoing project, aims at managing consent of users in order to regulate access to their personal data. In EnCoRe, a user is expected to define her preferences regarding consent, which are stored by enterprises. Once any piece of personal data is requested, these preferences are checked by the enterprises before granting access to the requested data. However, it may be cumbersome for users to define such complicated preferences. In our proposed solution, users' consent can be captured and managed dynamically by taking into account contextual information. Furthermore, our proposed approach offers more control and access to users as consent is stored and managed on their smartphone.

\section{A Case Study}
\label{sec:case_study}

In this section, we introduce the case study that we will use throughout the paper to demonstrate the feasibility of our approach. The case study is partially inspired from the European funded project ENDORSE\footnote{\url{https://ict-endorse.eu/}}. ENDORSE focuses on developing IT solutions for privacy preserving data management. An important aspect in ENDORSE is that of \emph{consent}. In the following, first we are going to provide the legal background in EU legislation about consent followed by more details about the capturing and managing of consent in healthcare scenarios.

\subsection{EU Legal Framework for Consent}
In this section, we present the EU directives to control access to personal data. In the following, we use the term \emph{data subject} to describe an individual whose data is handled, and \emph{data controller} to indicate any party that handles personal data. According to article 2(h) of the EU Data Protection Direction (DPD) \cite{EC1995}, consent is defined as: \emph{"the data subject's consent shall mean any freely given specific and informed indication of his wishes by which a data subject signifies his agreement to personal data relating to him being processed"}. The concept of consent enables a data subject to control access to her personal data. Furthermore, according to article 7 (a) of the EU DPD \cite{EC1995}, a data subject's personal data may only be processed if she has given her consent. Last but not least, data subjects may withdraw their consent at any time \cite{PDPA1999}.

In traditional healthcare systems, a data subject provides her paper-based consent typically once she is enrolled within the system. Generally, the paper-based consent is considered valid once signed by the data subject. Unfortunately, there are two main problems with the paper-based consent. First, it becomes very cumbersome for the data subject to withdraw her paper-based consent. That is, she has to go through complicated bureaucratic processes where she has to call on the responsible authority to withdraw her consent (most probably, as it is the case in Italy) with some considerable effort, waste of time, and a huge sense of frustration. Second, a data subject provides her consent in advance for all her medical data at the time of registration with the healthcare system even when it may not be necessarily used, thus violating the principle of least privilege.

In current IT healthcare systems, the notion of consent is captured as \emph{authorisation policies} that control the access to the data, such as in \cite{Russello2008}. Technically, the creation or editing of these authorisation policies is delegated to an IT security administrator. The security administrator operates on behalf of the data subject to deploy policies in the IT infrastructure of the data controller. In some countries, specific legislation may require the digital consent to be digitally signed by the data subject to be considered equivalent to the manually signed paper-based consent \cite{EC1999}.
\subsection{Healthcare Scenarios}

In this section, we describe several scenarios based on the IT healthcare system currently deployed in one of the major hospitals in Italy. A patient is provided with a smartphone where she can receive requests for giving her consent when she is interacting with the medical personnel. A patient can review through her smartphone who is requesting the access, the purpose of the request, and which data is requested.

At the time of providing consent, a patient may decide to save her preferences for subsequent consent requests made in the same context and/or by the same entity. Afterwards, a patient may withdraw her saved preferences regarding consent. Furthermore, a patient may activate withdrawn preferences regarding her consent. Last but not least, a patient may intend to delete, forever, her saved preferences for providing consent automatically.

\textbf{Patient visiting her GP.}
Let us consider the healthcare scenario where Alice moves to Milan and visits her GP for the first time. The GP requires access to Alice's medical history consisting of several medical tests and reports. For this purpose, the GP requires Alice's consent. Alice receives the consent request on her smartphone and decides to provide her consent also in the future.

\textbf{Patient visiting a cardiologist.}
Later, the GP of Alice discovers that she has a heart disorder. In this case, the GP refers Alice to a cardiologist for further testing. For visiting the cardiologist, Alice needs to contact the hospital booking service for getting an appointment. The hospital has several cardiologists thus it is not known in advance which one is assigned prior to the actual appointment. On the day of appointment, Alice will know the assigned cardiologist and can consent the cardiologist to access her medical data. However, Alice's consent should be valid for the duration of the treatment and the data accessed should be within the scope of the treatment (i.e., the cardiologist should not have access to Alice's gynecological reports). Moreover, if Alice is not happy with the assigned cardiologist then she may withdraw her consent and request a new cardiologist.

\textbf{Patient in an emergency situation.}
While Alice is driving in her car, she has a car accident and gets injured. The emergency response team reaches the accident location and starts treating Alice. For the treatment, the paramedic requires Alice's consent to access her medical history to get information about her allergies and any serious conditions that she already may have. Alice provides consent to access her medical records so that the paramedic is aware of her heart problem and provides the appropriate treatment that does not interfere with the treatment prescribed by the cardiologist. Although the paramedic has access to Alice's full medical record, consent should be revoked when the emergency is over.

\section{Overview of Teleo-Reactive Policies}
\label{sec:tr-ovewview}
From the above scenarios, it is clear that to capture all the details required to express the data subject's consent in different settings is very complex. If these details are not captured correctly by the security administrator in the policy specification then serious consequences might happen. In our experience, capturing all the security requirements through the specification of several independent authorisation policies is a very hard task. In the specific case of capturing a data subject's consent, it becomes even more complicated since there is the involvement of a human (which is the data subject that can grant, hold, and withdraw consent) and contextual information expressed in the policies (such as the location and time of the access).

In this paper, we propose to employ a goal-driven approach to \emph{glue} together and manage authorisation policies that have a common aim, that is the handling of consent. In particular, our observation is that we can simplify the specification of authorisation policies when these are treated as a \emph{program sequence} towards a specific goal. In this paper, we propose to leverage the idea of TR programs to glue together authorisation policies aiming at a specific goal. The idea of TR programs was initially introduced by Nilsson \cite{Nilsson1994}. A TR program is a control sequence directing towards a goal while taking into account changes in environmental circumstances. TR programs were used for automating behavioural robotics where a robot was continuously observing its environmental changes.

In the following, we provide a brief overview of TR policies that is similar to that introduced by Marinovic \emph{et al.} in \cite{Marinovic2010}.

\begin{figure} [htp]
\Keywords
\begin{lstlisting}[style=AMMA,breaklines,mathescape,rulesepcolor=\color{black}]
tr-policy $\mathit{name} (P_1, P_2, \ldots, P_m)$
$\mathit{cond}_1(V) \rightarrow \mathit{action}_1(V)$
$\mathit{cond}_{2_a} \wedge (\mathit{cond}_{2_b} \vee \neg \mathit{cond}_{2_c}) \rightarrow \mathit{action}_{2_y} \otimes \mathit{action}_{2_z}$
$\mathit{cond}_{3}(P_1) \rightarrow \mathit{action}_{3_a} \parallel \mathit{action}_{3_b}$
$\ldots$
$\mathit{cond}_{n_1} \wedge \mathit{cond}_{n_2} \ldots \vee \mathit{cond}_{n_x} \rightarrow \mathit{action}_{n_1} \parallel \mathit{action}_{n_2} \ldots \otimes \mathit{action}_{n_y}$
\end{lstlisting}
\caption{A TR Policy}
\label{fig:tr-policy}
\end{figure}

\subsection{TR Policy Representation}
A TR policy is an ordered list of rules as shown in Figure \ref{fig:tr-policy}, where each rule contains (Line 2) a condition part and an action part. The condition part contains a predicate that is bound with a variable, which is denoted with $V$. These variables may describe facts or states of the system or environment in which a TR policy is evaluated. A variable starts with a capital letter while a condition or an action starts with a small letter. The action part contains a function that is called by the TR policy. The action part may contain variables. The condition and action parts are separated by $\rightarrow$. Each TR-policy has a name starting with a small letter and can be instantiated with some parameters (Line 1). The condition part may include parameters, each denoted by $P_i$ (Line 4). The condition part can contain either a single condition or form (Line 2 or Line 6) a conditional expression where multiple conditions can be combined using logical operators $\wedge$ and $\vee$. Similarly, the action part can contain either a single function or multiple functions that may be executed sequentially and/or concurrently. The sequential and concurrent execution of functions can be represented with $\otimes$ operator (Line 3 and Line 6) and $\parallel$ operator (Line 4 and Line 6), respectively. In a TR policy, rules are specified in the descending order with respect to their priorities. That is, a high priority rule comes first.

\subsection{TR Policy Evaluation}
The runtime of the TR policy monitors changes in facts or states about the system or environment in which evaluation is performed. These changes can result in the condition part of a rule becoming either \emph{true} or \emph{false}. The functions in the action part of a rule will be executed if its condition part is evaluated to \emph{true} by the runtime. In a TR policy, the condition part corresponding to the highest priority rule is evaluated first. If it evaluates to \emph{false}, the condition part of the next high priority rule will be evaluated. In other words, if the action part of any rule is being executed, it means the condition parts of all higher priority rules (as compared to the current rule) are evaluated to \emph{false}. The action part of any rule is executed as long as its condition part evaluates to \emph{true} while condition parts of all higher priority rules (as compared to the current rule) remain \emph{false}.

\section{The ACTORS Approach}
\label{sec:proposed_approach}

ACTORS aims at automating creation and management of authorisation policies using a goal-driven approach. Authorisation policies are created and managed based on the users' intent while taking into account contextual information. The contextual information may be information about facts or states of the environment or the system. For collecting contextual information in an automated manner, we assume that users have smartphones equipped with some sensors for capturing environmental conditions. For instance, a smartphone can detect a fire alarm or an emergency situation such as a road accident.

ACTORS is based on three main parts including \emph{authorisation policies}, \emph{policy templates} and \emph{TR policies}. The main idea is that each TR policy captures a specific goal, such as managing consent for the GP. TR policies are used for instantiating authorisation policies from policy templates. TR policies also manage the lifecycle of instantiated authorisation policies. All three parts of ACTORS are managed by the user's smartphone.

Since all the details required in an authorisation policy may not be known in advance (such as, ID of the specific cardiologist assigned on the day of the visit, location where the visit will take place), we use policy templates to define abstract authorisation policies. When all the required information is available, TR policies can instantiate the required authorisation policies from the given templates. This instantiated authorisation policy is stored and enforced by the smartphone owned by the user, thus providing greater control to users to manage their consent.

\subsection{Authorisation Policies}
An authorisation policy specifies who is permitted (or denied) access to a resource under specific conditions. In ACTORS, an authorisation policy contains the following fields:

\begin{itemize}
	\item \textbf{Data Requester Role:} It is role of the entity who makes the access request. It can contain either a single role or a set of roles.
	\item \textbf{Data Requester ID:} It is ID of the one who makes the access request. Like the above field, this field can contain either a single ID or a list of IDs. This field is optional as permissions can be assigned to roles instead of specific IDs.
	\item \textbf{Data Subject ID:} It refers to the data subject who owns the resources.
	\item \textbf{Data Subject Resource:} It contains data subject resource(s) protected through the authorisation policy.
	\item \textbf{Access Rights:} Access rights define the permission on the data subject resource.
	\item \textbf{\emph{provided:}} It contains a conditional expression that may contain a set of conditions combined with $\mathit{and}$ and $\mathit{or}$ logical operators. Each condition is a predicate that is bound to a variable. These variables can come from contextual information that may be facts or states about the system or the environment. The contextual information may include access purpose, access time, access date, data requester location and data subject location.
\end{itemize}

\begin{figure} [htp]
\Keywords
\begin{lstlisting}[style=AMMA,breaklines,mathescape,rulesepcolor=\color{black}]
DataRequester.Role = {'Doctor' }
DataRequester.ID = {'Bob'}
DataSubject.ID = 'Alice'
DataSubject.Resource = {'Blood Test'}
AccessRights = {READ}
provided
		(AccessPurpose = 'Diagnosis' or
		AccessPurpose = 'Treatment') and
		AccessTime $\geq$ 9:00
\end{lstlisting}
\caption{An example of an authorisation policy}
\label{fig:example-authorisation-policy}
\end{figure}

Figure \ref{fig:example-authorisation-policy} illustrates an example of an authorisation policy where Bob in a role doctor is permitted to have read access on Alice's \emph{Blood Test} report provided he makes the access request after 9:00 hrs for the purpose of diagnosis or treatment. The use of the Data Requester ID might seem redundant given the fact that the policy already has a Data Requester Role. However, it might be the case that the data subject might not want a specific requester to access her data. For instance, Alice does not want Eve (another doctor and Bob's colleague) to read her \emph{Blood Test} report. This requirement can be captured by specifying in the Data Requester ID the condition \code{$\neg$ 'Eve'}.


\subsection{Policy Templates}
A policy template provides a structured format for instantiating authorisation policies on-the-fly. It is the authorisation policy specification with placeholders for variables which are assigned a value based on contextual information and a user's intent. A user's intent is about what a user can expect and can be captured based on actions taken by her. A policy template contains almost the same fields as an authorisation policy does. The fields of a very generic policy template are left blank so that they can be assigned a value based on contextual information. However, a list of options can be provided for each field. It means that a template field can only be filled, at the time of policy instantiation, with a value out of the list of options.

\begin{figure} [htp]
\Keywords
\begin{lstlisting}[style=AMMA,breaklines,mathescape,rulesepcolor=\color{black}]
DataRequester.Role = {'Dentist'}
DataRequester.ID
DataSubject.ID
DataSubject.Resource = {'Dental Report'}
AccessRights = {READ, WRITE}
provided
		AccessPurpose is 'Diagnosis' or 'Treatment'
\end{lstlisting}
\caption{An example of a policy template}
\label{fig:example-policy-template}
\end{figure}

Figure \ref{fig:example-policy-template} illustrates an example of a policy template. This policy template can be applied when a data requester is in role \emph{Dentist} and the requested resource is \emph{Dental Report} with access rights either \emph{READ} or \emph{WRITE} access and access purpose is either \emph{Diagnosis} or \emph{Treatment}. For rest of the fields, any value can be assigned based on contextual information and the user's intent.

Policy templates are associated with TR policies and goals that the TR policy is trying to achieve. For instance, the policy template in Figure \ref{fig:example-policy-template} can be applied when the goal of the patient is to visit a dentist. Therefore, such a template is associated with the TR policy managing that specific goal. Each TR policy can be associated with several templates. Based on contextual information and a user's intent, the TR policy can identify which policy template fulfils the criteria and then instantiates the required authorisation policy.

\subsection{TR Policies}
As already explained in Section \ref{sec:tr-ovewview}, TR programs were introduced for continuously monitoring the behaviour of a robot while taking into account environmental changes. In ACTORS, we use TR policies for controlling the lifecycle of authorisation policies towards a specific goal, which is the management of users' consent in a given situation. Each TR policy might be associated with several policy templates from which authorisation policies can be instantiated. Several TR policies might be present on the smartphone of the user. The selection of the appropriate TR policy is based on contextual information. The main advantage in using TR policies is that they provide a built-in prioritisation of actions needed for controlling the granting and revocation of users' consent that reacts to the changes in the context in which the users are interacting.

In the following section, we are going to provide details of how ACTORS can be used for the case study presented in Section \ref{sec:case_study}.

\section{Managing Consent in Healthcare Scenarios}
\label{sec:application}

ACTORS can be applied to any domain; however, we focus on healthcare scenarios as already described in Section \ref{sec:case_study}, where consent needs to be captured and saved based on contextual information and the patient's intent. For automatically instantiating authorisation policies regarding consent and managing lifecycle of those policies, we assume that each patient is provided a set of TR policies and policy templates at the time of registration with her healthcare provider. In fact, TR policies and policy templates are deployed on patients' smartphones together with an application. Each TR policy can be associated with multiple policy templates. The smartphone application automatically selects the most appropriate TR policy and the policy template based on the consent request and contextual information. After instantiation of authorisation policies regarding consent, they are stored and enforced by the patient's smartphone. It should be noted here that only policies and patient's decisions are stored in the smartphone while the medical data is stored in the caregiver IT infrastructure.  In this section, we explain in detail how we exploit the proposed approach, described in Section \ref{sec:proposed_approach}, for providing solutions for each scenario described in Section \ref{sec:case_study}.

\textbf{Patient visiting her GP.}
In the scenario when a GP needs the patient consent, a consent request is sent to the patient for providing access to a GP to requested resources. This consent request may be directly sent by the healthcare system to the patient when a GP makes an access request to the patient resources. This consent request may include information about the GP and the patient, the patient resources, an access purpose and access duration details. Based on the consent request together with contextual information, the most appropriate applicable TR policy and policy template are selected.

\begin{figure} [htp]
\Keywords
\begin{lstlisting}[style=AMMA,breaklines,mathescape,rulesepcolor=\color{black}]
tr-policy $\mathit{consentAtGPClinic(Patient)}$

$\mathit{consentAvailable(Patient, GP)}$  $\wedge$  $\mathit{saveCurrentPreferences}$  $\rightarrow$  $\mathit{instantiatePolicy(Patient)}$  $\otimes$  $\mathit{activate(Patient.Policy)}$ $\parallel$ $\mathit{sendConsent(Patient, GP)}$

$\mathit{consentAvailable(Patient, GP)}$  $\rightarrow$  $\mathit{sendConsent(Patient, GP)}$

$\mathit{needsConsent(Patient, GP)}$  $\wedge$  $\mathit{instantiatedPolicy(Patient)}$  $\wedge$  $\neg \mathit{withdrawn(Patient.Policy)}$  $\rightarrow$  $\mathit{evaluatePolicy(Patient)}$

$\mathit{needsConsent(Patient, GP)}$  $\rightarrow$  $\mathit{waitPatientDecision(Patient, GP)}$

$\mathit{deleteSavedPreferences(Patient)}$  $\rightarrow$  $\mathit{remove(Patient.Policy)}$

$\mathit{activatePolicyRequest(Patient)}$  $\rightarrow$  $\mathit{activate(Patient.Policy)}$

$\mathit{withdrawPolicyRequest(Patient)}$  $\rightarrow$  $\mathit{withdraw(Patient.Policy)}$

\end{lstlisting}
\caption{A TR policy for managing authorisation policy for providing consent to a GP}
\label{fig:gp-tr-policy}
\end{figure}

Figure \ref{fig:gp-tr-policy} describes a TR policy that is applied when a GP needs a patient's consent for accessing her data from his clinic. The name of this TR policy is $\mathit{consentAtGPClinic}$ whereas \emph{Patient} is the parameter. When the first consent request is made, consent is not available and the condition parts of rules at Line 3 and Line 5 evaluate to \emph{false}. The condition part of rule at Line 7 also evaluates to \emph{false} as no authorisation policy is instantiated yet i.e., $\mathit{instantiatedPolicy(Patient)}$ is \emph{false}. However, the condition part of rule at Line 9 evaluates to \emph{true}, so the action part of this rule is executed and the system waits for the patient decision for providing consent to her GP i.e., $\mathit{waitPatientDecision(Patient, GP)}$ is executed.

Once the patient provides consent for granting access to her GP on her resources, then $\mathit{consentAvailable(Patient, GP)}$ becomes \emph{true}. At the time of providing consent, a patient can be given an option to save her current preferences for providing her consent for similar consent requests when made in the same environment. If a patient does so, the condition part of rule at Line 3 becomes \emph{true}; therefore, the authorisation policy regarding consent is instantiated from the policy template and then it is activated while at the same time, consent is sent.

\begin{figure} [htp]
\Keywords
\begin{lstlisting}[style=AMMA,breaklines,mathescape,rulesepcolor=\color{black}]
DataRequester.Role = {'GP'}
DataRequester.Name
DataSubject.Name
DataSubject.Resource
AccessRights
provided
		AccessPurpose is 'Diagnosis' or 'Treatment'
		AccessTime is within DutyHours
		DataRequester.CurrentLocation = DataSubject.CurrentLocation
		DataRequester.CurrentLocation = DataRequester.Clinic.Location
\end{lstlisting}
\caption{A policy template for generating an authorisation policy for providing consent to a GP}
\label{fig:gp-policy-template}
\end{figure}

Figure \ref{fig:gp-policy-template} illustrates a policy template that is applied when a patient visits her GP, as is evident from the data requester role that is GP only. The empty fields including data requester name, data subject name, data subject resource and access rights can be filled with values based on the consent request. However, there are certain conditions in the \emph{provided} part of the policy template that are formulated at the time of instantiating an authorisation policy. These conditions include: the access purpose must be either \emph{diagnosis} or \emph{treatment}; access time must be in office hours; and both the patient and the GP must be present in the GP's clinic. These conditions are formulated based on contextual information that is collected from either patient's smartphone or the external information point, such as made available by the healthcare provider. The contextual information from a patient's smartphone may include information like patient's current location, while contextual information from the external information point may include information about location of GP's clinic and GP's duty hours. Once all the required information for the applicable policy template is retrieved, the authorisation policy is instantiated and activated.

\begin{figure} [htp]
\Keywords
\begin{lstlisting}[style=AMMA,breaklines,mathescape,rulesepcolor=\color{black}]
DataRequester.Role = {'GP'}
DataRequester.ID = {'Bob'}
DataSubject.ID = 'Alice'
DataSubject.Resource = {'Blood Test'}
AccessRights = {READ}
provided
		AccessPurpose = 'Diagnosis' and
		(AccessTime $\geq$ 9:00 and AccessTime $\leq$ 17:00) and
		DataSubject.CurrentLocation = 'Milan' and
		DataRequester.CurrentLocation = 'Milan'
\end{lstlisting}
\caption{An authorisation policy for providing consent to a GP}
\label{fig:gp-authorisation-policy}
\end{figure}

Figure \ref{fig:gp-authorisation-policy} shows the instantiated authorisation policy regarding consent, expressing that a GP Bob can get patient Alice's consent for \emph{READ} access on Alice's \emph{Blood Test} when accessed for the \emph{Diagnosis} purpose during the duty hours (that is, between 9:00 and 17:00 hrs) from Bob's clinic located in \emph{Milan}.

A patient may decide to withdraw her consent. In this case, the condition part of rule at Line 15, i.e. condition $\mathit{withdrawPolicyRequest(Patient)}$, becomes \emph{true} and the authorisation policy is withdrawn by invoking $\mathit{withdraw(Patient.Policy)}$ function. Furthermore, a patient can decide to activate her withdrawn consent. In this case, condition $\mathit{activatePolicyRequest(Patient)}$ becomes \emph{true} and $\mathit{activate(Patient.Policy)}$ function is invoked for activating the authorisation policy. Last but not least, a patient may also choose to delete forever her saved preferences for automatically providing consent. In this case, $\mathit{deleteSavedPreferences(Patient)}$ becomes \emph{true} and $\mathit{remove(Patient.Policy)}$ function is invoked for deleting the instantiated authorisation policy.

In case if a GP needs the patient consent when the patient has already saved preferences for providing consent automatically to her GP and consent is not withdrawn yet then consent will be provided after evaluating the consent request and contextual information against the instantiated authorisation policy, see rule in Figure \ref{fig:gp-tr-policy} at Line 7. We assume that the consent request is same as already described above. However, we have to collect contextual information in order to evaluate the authorisation policy for providing consent. The patient's smartphone may provide information about her location and the current time while the information about the GP's location can be collected from the external information point. This may be the healthcare system or the GP's smartphone which may provide GP's location information to the patient's smartphone. Based on the consent request and contextual information, the authorisation policy is evaluated (see rule in Figure \ref{fig:gp-tr-policy} at Line 7). After the evaluation of the authorisation policy, consent becomes available and the consent response is automatically sent by the patient's smartphone (see rule in Figure \ref{fig:gp-tr-policy} at Line 5). The consent response contains patient consent if the authorisation policy evaluates to \emph{true}, otherwise it may contain an error message.

A patient may decide not to save her current preferences for providing consent automatically to her GP. In such a case, the patient will be explicitly asked each time (see rule in Figure \ref{fig:gp-tr-policy} at Line 9) and consent will be provided once the patient takes her decision (see rule in Figure \ref{fig:gp-tr-policy} at Line 5).

\textbf{Patient visiting a cardiologist.}
A cardiologist may also need the patient consent while accessing the patient resources. Like the above scenario, a patient receives the consent request. This consent request may include information about the cardiologist and the patient, the patient resources, an access purpose and access duration details. The additional point in this scenario as compared to the previous scenario is that a cardiologist is provided consent for getting access on the patient resources as long as the treatment may last. In other words, the saved preferences for providing consent are deleted automatically right after the treatment.

\begin{figure} [htp]
\Keywords
\begin{lstlisting}[style=AMMA,breaklines,mathescape,rulesepcolor=\color{black}]
tr-policy $\mathit{consentAtSpecialistClinic(Patient)}$

$\mathit{consentAvailable(Patient, Specialist)}$  $\wedge$  $\mathit{saveCurrentPreferences}$  $\rightarrow$  $\mathit{instantiatePolicy(Patient)}$  $\otimes$  $\mathit{activate(Patient.Policy)}$ $\parallel$ $\mathit{sendConsent(Patient, Specialist)}$

$\mathit{consentAvailable(Patient, Specialist)}$  $\rightarrow$  $\mathit{sendConsent(Patient, Specialist)}$

$\mathit{needsConsent(Patient, Specialist)}$  $\wedge$  $\mathit{instantiatedPolicy(Patient)}$  $\wedge$  $\neg \mathit{withdrawn(Patient.Policy)}$  $\rightarrow$  $\mathit{evaluatePolicy(Patient)}$

$\mathit{needsConsent(Patient, Specialist)}$  $\rightarrow$  $\mathit{waitPatientDecision(Patient, Specialist)}$

$\mathit{timeout(Patient.Policy)}$  $ \vee$  $\mathit{deleteSavedPreferences(Patient)}$  $\rightarrow$  $\mathit{remove(Patient.Policy)}$

$\mathit{activatePolicyRequest(Patient)}$  $\rightarrow$  $\mathit{activate(Patient.Policy)}$

$\mathit{withdrawPolicyRequest(Patient)}$  $\rightarrow$  $\mathit{withdraw(Patient.Policy)}$

\end{lstlisting}
\caption{A TR policy for providing consent to a specialist}
\label{fig:cardiologist-tr-policy}
\end{figure}

\begin{figure} [htp]
\Keywords
\begin{lstlisting}[style=AMMA,breaklines,mathescape,rulesepcolor=\color{black}]
DataRequester.Role = {'Cardiologist'}
DataRequester.ID
DataSubject.ID
DataSubject.Resource = {'ECG Report', 'Cardiography', 'Engyography'}
AccessRights = {READ, WRITE}
provided
		AccessPurpose is 'Diagnosis' or 'Treatment'
		AccessTime is within DutyHours
		DataRequester.CurrentLocation = DataSubject.CurrentLocation
		DataRequester.CurrentLocation = DataRequester.Clinic.Location
\end{lstlisting}
\caption{A policy template for generating an authorisation policy for providing consent to a cardiologist}
\label{fig:cardiologist-policy-template}
\end{figure}

Figure \ref{fig:cardiologist-tr-policy} shows the TR policy for managing authorisation policy in order to provide consent to a specialist. The name of this TR policy is $\mathit{consentAtSpecialistClinic}$. The TR policy is similar to one already described in Figure \ref{fig:gp-tr-policy}. In case of a cardiologist, the TR policy of specialist is selected. As we can observe that the TR policy of a specialist is very generic, it can be applied to other specialists such as a dentist and a gynaecologist. However, there is a specific policy template for each specialist. The policy template for cardiologist is shown in Figure \ref{fig:cardiologist-policy-template}. The policy template is restricted to only resources that could be accessed by a cardiologist. These resources include \emph{ECG Report}, \emph{Cardiography} and \emph{Engyography}. This is different from the policy template of above scenario as resource field in Figure \ref{fig:gp-policy-template} is left empty, indicating that a GP can obtain consent to access any resource.

\begin{figure} [htp]
\Keywords
\begin{lstlisting}[style=AMMA,breaklines,mathescape,rulesepcolor=\color{black}]
DataRequester.Role = {'Cardiologist'}
DataRequester.ID = {'David'}
DataSubject.ID = 'Alice'
DataSubject.Resource = {'ECG Report'}
AccessRights = {READ, WRITE}
provided
		AccessPurpose = 'Diagnosis' and
		(AccessTime $\geq$ 9:00 and AccessTime $\leq$ 17:00) and
		DataSubject.CurrentLocation = 'Como' and
		DataRequester.CurrentLocation = 'Como'
\end{lstlisting}
\caption{An authorisation policy for providing consent to a cardiologist}
\label{fig:cardiologist-authorisation-policy}
\end{figure}

Figure \ref{fig:cardiologist-authorisation-policy} shows the authorisation policy regarding consent for a cardiologist when a patient intends to save her preferences until she is treated. The authorisation policy expresses that a cardiologist David can get patient Alice's consent for \emph{READ} and \emph{WRITE} access on Alice's \emph{ECG Report} when accessed for \emph{Diagnosis} purpose during the duty hours (that is, between 9:00 and 17:00 hrs) from David's clinic located in \emph{Como}.

The authorisation policy regarding consent for a cardiologist may automatically be deleted once the treatment completes. This information about treatment duration can be collected by the patient at the time of saving her preferences. For instance, it may be included in the consent request or can be collected as contextual information from the information point made available by the service provider. Once the treatment duration expires (starting from when the first consent request is made), condition $\mathit{timeout(Patient.Policy)}$ becomes automatically \emph{true} and $\mathit{remove(Patient.Policy)}$ function is invoked for deleting the instantiated authorisation policy according to the rule at Line 11 in Figure \ref{fig:cardiologist-tr-policy}. Alternatively, a patient may decide to delete her saved preferences during the treatment duration as already considered in above scenario.

\textbf{Patient in an emergency situation.}
In an emergency situation, the emergency response team may need a patient's consent in order to get an access to her medical data for the treatment purpose. Similar to above scenarios, the patient receives the consent request, which may include information about the emergency response team, the patient resources, an access purpose and access duration details. Similar to the cardiologist scenario, we consider that the patient intends to provide her consent as long as the treatment may last. Technically, the saved preferences for providing consent are deleted automatically right after the treatment. The TR policy for specialist, shown in Figure \ref{fig:cardiologist-tr-policy}, can also be applied for this scenario.

\begin{figure} [htp]
\Keywords
\begin{lstlisting}[style=AMMA,breaklines,mathescape,rulesepcolor=\color{black}]
DataRequester.Role = {'EmergencyResponseTeam'}
DataRequester.Name
DataSubject.Name
DataSubject.Resource = {'Allergy Report', 'Blood Test'}
AccessRights = {READ}
provided
    There is an Emergency situation
		AccessPurpose is 'Diagnosis' or 'Treatment'
		DataRequester.CurrentLocation = DataSubject.CurrentLocation
\end{lstlisting}
\caption{A policy template for generating an authorisation policy for providing consent to the emergency response team}
\label{fig:es-policy-template}
\end{figure}

The policy template applied in emergency situation is shown in Figure \ref{fig:es-policy-template}. In the \emph{provided} part of the policy template for emergency situations, we include the condition for capturing the notion of emergency situation, i.e., \emph{There is an Emergency situation}. Furthermore, we omit also the condition \emph{AccessTime is within DutyHours}, in contrast to the policy template for a GP shown in Figure \ref{fig:gp-policy-template}, considering the fact that the emergency can happen at any time. For restraining access in emergency situations, the resource field of the policy template is set to \emph{Allergy Report} and \emph{Blood Test}. Moreover, we consider \emph{READ} only access in emergency situations.

\begin{figure} [htp]
\Keywords
\begin{lstlisting}[style=AMMA,breaklines,mathescape,rulesepcolor=\color{black}]
DataRequester.Role = {'EmergencyResponseTeam'}
DataRequester.ID = {'Fayne'}
DataSubject.ID = 'Alice'
DataSubject.Resource = {'Allergy Report'}
AccessRights = {READ}
provided
		Emergency = TRUE and
		AccessPurpose = 'Diagnosis' and
		DataSubject.CurrentLocation = 'Aachen' and
		DataRequester.CurrentLocation = 'Aachen'
\end{lstlisting}
\caption{An authorisation policy for providing consent to the emergency response team}
\label{fig:es-authorisation-policy}
\end{figure}

Figure \ref{fig:es-authorisation-policy} shows the authorisation policy for providing consent to the emergency response team. This authorisation policy is instantiated when an emergency happens in \emph{Aachen} and Fayne, a member of emergency response team, requests \emph{READ} access on (a patient) Alice's \emph{Allergy Report} for \emph{diagnosis} while Alice provides her consent and also saves her preferences for subsequents requests in the same environment. The occurrence of emergency situation may be detected using a patient's smartphone.

There are few important points to be considered. First, we are instantiating one authorisation policy per instance of the emergency response team. Alternatively, it may also be possible to instantiate the authorisation policy at the role level (i.e., EmergencyResponseTeam) instead of at the instance level (i.e., Fayne). Second, the patient may be in the unconscious state and may not be able to provider her consent. In such situations, authorisation policies can be instantiated from break-the-glass policy templates without asking patients. In other words, the emergency response team may provide consent on patient's behalf when patients are in the unconscious state. Here, the unconsciousness state can be incorporated at the time of sending consent request by members of the emergency response team to the patient's smartphone. Finally, as ultimate break-the-glass in case the smartphone is not reachable or not functioning, the emergency team can specify the current circumstances together with the request for accessing the patient's medical data. This information then can be checked in a post-incident analysis to make sure that such access mode is not abused. Again, it should be noted here that the medical data are not stored in the smartphone.

\section{Conclusions and Future Work}
\label{sec:conclusion_future_work}
With the increasing attention towards the notion of data subjects consent to be integrated in access control mechanisms, the task of properly capturing security requirements in policy specification is becoming very daunting. This increases the risk of introducing errors in the policy specification that might compromise the privacy of the medical data. In the light of this, in this paper we have proposed ACTORS a goal-driven approach, where authorisation policies are managed by TR policies that have goal of capturing the consent preferences of a data subject. As we have shown in our scenario, data subjects might want to handle consent in accordance with the actual situation and context. TR policies are structured in such a way that rules at the top are closer to the goal of the policy while rules at the bottom are more relevant when the goal is not close to be achieved. This is very natural for humans to grasp; therefore, a security administrator can capture more naturally the security requirements.

As future work, we are planning to perform a thorough evaluation in the ENDORSE project. Our current experience so far with the capturing of security requirements with TR policies is very promising. We have so far captured the requirements of one of the testbeds in the project, which is the medical scenario. In the coming months, we are going to evaluate in a second testbed mainly focused on capturing consent for handling personal data of customers of a commercial entity.

Another area that we want to investigate is that of enforcing cross-domain policies. In this setting, it is difficult for the security administrator to have all the details of the different domains in which the data of the user might end up. Our idea is to have mapping of the policy templates from one domain to the other. Currently, we are planning to perform the mapping of the policy templates by means of ontologies.

\section*{Acknowledgment}
This work is supported by the EU FP7 programme, Research Grant 257063 (project ENDORSE).

\bibliographystyle{IEEEtran}
\bibliography{IEEEabrv,references}


\end{document}